\documentstyle[preprint,aps]{revtex}

\input epsf

\begin{document}
\draft
\preprint{}

\def\r{{\bf r}} \def\q{{\bf q}}

\title{Solid like friction of a polymer chain}

\author{Thierry Charitat and Jean-Fran\c{c}ois Joanny}
\address{Institut Charles Sadron, 6 rue Boussingault 67083 Strasbourg
Cedex France\\Email:charitat@ics.u-strasbg.fr}
\date{\today}
\maketitle
\begin{abstract}

We propose a simple friction model for isolated polymer chains
on a solid substrate. The chains are
pulled at constant velocity by one end, the other end can
be trapped on the solid substrate on localised sites. We
focus on the energy dissipation due to the traps. This simple model leads
to non trivial friction laws, depending
on the velocity and the distance between traps. Some refinements of the
model such as the effect of thermal fluctuations are also reported.

\end{abstract}

%
%

\pacs{61.41.+e: polymer, elastomer, and plastics.\\
68.45.-v: solid-fluid interfaces\\
81.40.Pq: friction, lubrification, and wear.\\
83.10.Nn: Polymer dynamics.\\
submitted to {\em Eur. Phys. J. E}}

%
%

\narrowtext

\vspace{1cm}

Friction phenomena occur at the interface between two media under
relative motion. Their understanding is crucial for many technological
applications and raises many fundamental questions. This is one of the
oldest
problems in physics but there is still a relatively poor understanding of the
microscopic mechanisms
of friction.

Dry solid friction is the older known form of friction, first
introduced by da
Vinci and Amonton. The dry friction laws give the friction force in terms
of the
load and the static
and dynamic friction coefficients $\mu_s$ and $\mu_d$. A first surprising
result is that
the friction coefficients do not depend of the contact area. The first
microscopic
model of Amonton's law \cite{tabor,greenwood} appeared in the middle of
this century and shows the importance of the interface
roughness.
The dynamic friction coefficient depends only very weakly on the sliding
velocity $u$ \cite{persson,baumberger1,caroli,baumberger2}, and in practice,
the friction force reaches
a constant value at vanishing velocity. There is still
a poor
understanding of dry friction, both from the theoretical
and the experimental points
of view, in part due to the
difficulties to control the interfacial properties \cite{persson}.

Fluid friction can be studied within the framework of hydrodynamics
and is better understood. For a solid moving in a newtonian fluid,
the friction law  involves the
viscosity of the
fluid $\eta$, its density $\rho$, its size $R$ and the velocity $u$. The
dimensionless
Reynolds number $R_e = \frac{\rho u R}{\eta}$ gives the ratio between
the inertial and
viscous forces. A simple scaling argument leads to the force acting on the
solid as:
$f \sim \eta R u  g(R_e)$, where $g$ is a function of the Reynolds
number.
At low velocities, ie low Reynolds numbers, the friction force must be
independent of the density $\rho$,
and thus is proportional to $\eta R u $,
(Stokes
friction). At high Reynolds number, the viscous stresses are negligible and
$f \sim \eta R u R_e \sim \rho R^2 u^2$.

The friction between two polymeric media is a subject of major interest
for many
practical applications (lubrification, adhesion, extrusion, wetting).
Polymeric systems also allow to work on well
controlled
surfaces and several microscopic parameters that can
influence friction can be varied, such as  a chain
grafting density, chain
lengths, chain affinity ... On the other hand, the complexity of the
dynamic
of the interfacial
chains leads to a large variety of frictional behaviors: gel \cite{osada,gong,casoli}
or melt friction, slip transition \cite{migler,durliat,gay1} and stick-slip 
\cite{mhetar,koustos}. 

In a recent paper \cite{subbotin}, Subbotin {\it et al} proposed a model 
describing the friction of confined polymers melt. In their model, the energy dissipation is 
simply due to the high
friction of monomer anchored on the surface and the effect of 
loops and bridges on the friction law were precisely studied. In this paper, we 
present a simple friction model 
for a single polymer chain on a solid
surface, focusing mainly on the microscopic mechanism of energy dissipation. We consider 
a polymer chain in a solvent, pulled by one of its extremities (s=N where s is the 
monomer number) at a constant
velocity $u$ (figure \ref{generale}).
The other extremity (s=0) can be trapped on the surface by localised
defects. Our aim is to calculate the dissipated energy due to the
traps. This extremely simple and naive model gives a non trivial behavior
of the friction force which is similar to solid friction in the sense
that it may not vanish in the limit of small velocities. It could be also
possible to test this behavior by micromanipulation of isolated polymer chains in a
solvent
with an atomic force microscope. Finally, it could be a
possible way to introduce an energy dissipation in molecular motor models
\cite{prost}.

We first present the one chain problem on a linear and periodic array of
traps. In this first
part, we simply assume that the trapped monomer is leaving the trap when
the tension reaches a critical value $\tau$. In the second part, we show
how thermal
fluctuations can induce the desorption of this monomer trapped in a potential
barrier.

In a first approximation, the dynamics of a polymer chain can be described
by the well-known Rouse model \cite{doi},
which neglects
hydrodynamic interactions. The polymer chain is represented as a set of
N beads  at position $z$ connected
by springs with entropic rigidity $k = \frac{k_B T}{a^2}$, where $a$ is the
monomer size.
Each monomer behaves as a solid particle with a friction coefficient
$\zeta$ proportional
to the solvent viscosity $\eta$, $\zeta \simeq \eta a$. The Rouse
equation which describes the
dynamics of the chain is written(in one dimension) as:

\begin{equation}
\zeta \frac{\partial z(s,t)}{\partial t}=k \frac{\partial^2}{\partial
s^2}z(s,t) +f(t)
\label{rouse0}
\end{equation}

where $z(s,t)$ is the position of monomer $s$ at time $t$ and $f(t)$ 
is a Langevin random force.
The  slowest relaxation mode of the chain has a characteristic time
(the Rouse time) $T_R=\frac{\zeta N^2}{k}$.
As it ignores hydrodynamic interactions, the Rouse model
does not give a good description of
polymers in dilute solutions. Nevertheless, the Rouse description is
conceptually important
and we choose it as a starting point to describe our results on polymer
friction.
In appendix A we show how
it is possible to take into account hydrodynamic interactions within the
framework of the Zimm model.
This complication does not change qualitatively the behavior of the
friction force. Our results are obtained for gaussian chains in a $\Theta$ 
solvent. They can easily be generalised to polymer in a good solvent, but 
for a sake of clarity we do not show the results in the present paper.

Our paper is organised as follows. In the first section, we consider
the interaction between one chain and a linear array of traps along
the direction of the velocity, ignoring any thermal fluctuations. The traps are
pointlike on the surface and the chain end point remains trapped
as long as the chain tension is smaller than a critical value
$\tau$ which is the maximum force that can be exerted on the chain end 
by the trap. We first discuss the single trap problem and then a
periodic one-dimensional array of traps with a period $d$. In
appendix D we relate this  model to the more realistic problem where
the surface has a finite density of traps $n$ per unit area. Complete
calculations can be performed within the Rouse model; in the main
text of the paper, we rather give simple scaling arguments postponing
the detailed calculations in an appendix. Section \ref{fluc} is
devoted to the effect of thermal fluctuations. We describe each trap
more realistically as a potential well and we estimate the escape
probability of the trapped chain end when the tension increases.
This amounts to calculating an effective critical tension with which
the results of the previous section can be used. Following reference
\cite{seifert}, we describe the release of the chain end as a Kramer
process and we use a self-consistent force approximation.
The last section summarises our results and discusses some possible
extensions.

\section{Polymer chain interacting with traps}

\subsection{The one trap problem.}

We study a Gaussian polymer chain interacting with one trap on a solid
surface. For simplicity, we consider the chain as two-dimensional and
lying in the surface with a Gaussian conformation (at
rest). The end point of the chain $s=N$
moves at a velocity $u$.
When the chain is free (not trapped), its average conformation can be
obtained from the Rouse equation \ref{rouse0}. The average position
$z_{free}(s,t)$ of monomer $s$ at time $t$ is

\begin{equation}
z_{free}(s,t)= \frac{1}{2}\frac{\zeta u}{k}s^2 = \frac{1}{2}u T_R
\left(\frac sN \right)^2
\label{freesol}
\end{equation}

The tension of the chain at monomer $s$ is given by the friction force on
the chain end section containing $s$ monomers
$\tau_{ch}(s,t)= k \frac{\partial z_{free}(s,t)}{\partial s} = \zeta u s$. The
tension on the
free end of the chain vanishes, and the external force applied to pull the
chain is
$N\zeta u = \frac kN u T_R$. This can be viewed either as the friction force on
the N monomers, or as the
elastic tension of N springs in series extended over a distance $u T_R$.
In the absence of trapping, the energy dissipated per unit time by the solvent 
viscosity is simply $\frac{d {\cal E}_d}{dt} = N\zeta u^2$.\\

When the end monomer (s=0) gets trapped, the chain is progressively
elongated by the
motion of the pulled end (s=N). The tension applied on the trapped monomer
increases until it
reaches the critical value $\tau$. The trapped chain end is then
released and the chain contracts back to its free conformation.

The Rouse equation for the trapped chain
allows the calculation of the chain tension (appendix \ref{A})~:

\begin{equation}
\tau_{ch}(s,t) = k \frac{u T_R}{N} \left[ \frac{t}{T_R} + \frac sN
+\frac{2}{\pi^2}
\sum_{n=1}^{+\infty}\frac{1}{n^2} \left( 1 - \exp{(-n^2\pi^2\frac{t}{T_R})}
\right) \cos{\left( \pi \frac{n}{N}(s-N) \right)} \right]
\label{tenstrapped}
\end{equation}

 The tensions on the trapped end 
$\left( \tau_{tr} \left( t \right) = \tau_{ch} \left( s=0,t \right) \right)$ and on the pulled 
end $\left( \tau_{ch} \left( s=N,t \right) \right)$ are plotted on
figure \ref{tension}. At short time after the trapping ($t \ll
T_R$), the tension on the pulled end
remains constant (equal to the viscous friction force $N \zeta u$), and the tension
on the trapped monomer increases as 
$\tau_{tr} \sim \zeta N u \left(\frac{t}{T_R}\right)^{1/2}$. 
At larger times, ($t \gg T_R$), the tension is uniform along the chain and
both tensions increase as $\zeta N u \frac{t}{T_R}$. After the sticking time
$t_s$, the tension $\tau_{tr}$ becomes equal to the critical value $\tau$.

The characteristic velocity of the problem is obtained by comparing
the friction force on the chain $\zeta N u$ to the maximum force $\tau$
that can be exerted by the trap
 $u_c = \frac{\tau}{N\zeta}= \frac{N\tau}{k T_R}$.

If $u \ll u_c$, the viscous force on the chain is smaller than $\tau$. When the
trapped end is released from the trap, the whole chain is stretched with a 
constant tension and its extension is $L =\frac{N\tau}{k}$. The
tension acting on the first monomer at time $t$ is $\tau_{tr}(t) \simeq
\frac{k}{N}ut$ and
the sticking time $t_s$ is given by~:

\begin{eqnarray}
\nonumber
t_s &=& \frac{u_c}{u} T_R  \hspace{2.5cm} (u \ll u_c \Leftrightarrow t_s
\gg T_R)
\label{td2}
\end{eqnarray}

The work performed to extend the chain is $\frac{N \tau^2}{2k}$.
After the release of the trapped end, the elastic energy stored in the 
chain $W_d = \frac{N\tau^2}{2k}$ is
dissipated by a viscous process. This argument is essentially
identical to the classical argument of Lake and Thomas \cite{lake}
which has been used to study the adhesive properties of polymers by
de Gennes \cite{pggl,pgg2}.

If $u \gg u_c$, the viscous force on the chain is larger than
$\tau$, and only a small part of the chain involving $s_0 (t) \sim
\sqrt{\frac{k}{\zeta} t}$ monomers is elongated when the
release of the trapped end occurs. It contributes to the tension on the
trapped monomer as
$\tau_{tr} (t) \simeq \frac{k}{s_0 (t)}ut$ and the sticking time $t_s$ is
now given by~:

\begin{eqnarray}
\nonumber
t_s &=& \left( \frac{\sqrt{\pi}}{2}\frac{u_c}{u} \right)^2 T_R= \frac{\pi
\tau^{2}}{4ku^{2}\zeta}
\label{td1}
\end{eqnarray}
where the numerical prefactor is obtained from the detailed
calculations of the appendix. The force acting on the pulled monomer remains 
constant and  equal to $N \zeta u $ during
the trapping with a good approximation. After the release of the traped end, the 
tension of the pulled
monomer is calculated in appendix \ref{B} and is given
by (\ref{tensrelaxed}). The extra energy dissipated due to the trapping is of the
order of the elastic energy stored during the trapping in the
deformed part of the chain
$W_{d}=s_{0}(t_s)\frac {\tau^{2}}{k}=N \frac{\tau^2}{k} {\frac{u_{c}}{u}}$. More 
complete calculations confirming these scaling law can be
done using the Rouse model. They are presented in appendices \ref{A}
and \ref{B}. After the
release of the anchored monomer, the chain relaxes to the steady state,
and its elastic
energy is dissipated. The total dissipated energy is

\begin{equation}
W_d = N \zeta u^2 t + \frac{4}{3\pi^{1/2}} N\zeta u^{2}t_s
\left(\frac{t_s}{T_R}\right)^{1/2}
\label{work}
\end{equation}

\subsection{The many-traps problem.}

We now consider a polymer pulled at a velocity u on a two dimensional
surface with a
density of traps n. We refer to appendix \ref{D} for a discussion of the
trapping
probability that gives the mean distance $d(u)$ between trapping events.
The problem
is then equivalent to a one dimensional array of traps along the velocity
direction with
an average density $1/d(u)$. For simplicity, we consider here a periodic
lattice with a
distance d(u) between traps. We then calculate
the average friction force in a steady state as $F_{fr}=\frac {W_{d}}{d}$
where
$W_{d}$ is the energy dissipated on each trap.

We first consider the limit of very low velocities $u\ll u_{c}$. When
the chain is trapped on one defect and is just on the verge of
detaching, the chain has a constant tension $\tau$
and its size is $L=\frac{N\tau}{k}$. If this is smaller than the
distance $d$ between traps, the traps are independent and we can
use the results of the previous section. The energy dissipated per
trap is $W_{d} \sim \frac{N\tau^{2}}{ k}$ and the friction force is
$F_{fr} \sim \frac{N\tau^{2}}{kd}$. If $d\ll N\tau/k$, the chain end gets 
trapped before the full
relaxation of the tension. The number of monomers where the tension
has relaxed increases as $t^{1/2}$ and the position of the first
monomer is given by $z(t)\sim \frac{\tau t^{1/2}}{(\zeta k)^{1/2}}$. The
chain gets trapped again when $z(t)=d$ or after a time $t_{a} \sim \frac
{\pi \zeta
kd^{2}}{4 \tau^{2}}$. The tension then becomes uniform along the chain over
the Rouse time $T_{R}$ and equal to $\tau'=k(L-d)/N$. The
chain stretches then back to a uniform tension $\tau$ over the
time $d/u$. The dissipated energy per trap is
$W_{d}=\frac{N\tau^{2}}{2k}-\frac{N\tau'^{2}}{2k}$ and the friction
force is $F_{fr} \sim \tau$.

This quasi-static description remains valid as long as $d/u>T_{R}$. If
this is not the case, the tension does not have time to relax to the
smaller value $\tau'$ over the whole chain and remains equal to
$\tau$ (in fact sightly larger) on the pulled monomer. As long as the
viscous force is small ($u<u_{c}$) this tension balances the friction
force and thus $F_{fr}\sim \tau$. In this case, the dissipated
energy per trap is of the order of the elastic energy of a chain
spanning the distance $d$ i.e. containing $s=kd/\tau$ monomers.

If the velocity $u$ is larger than the critical value $u_{c}$, the
total dissipation is dominated by the viscous forces. The defects are
independent if $t_s (u) \ll \frac du$ i.e. if $d>\frac{\tau^{2}}{k
\zeta u}$. In this case the dissipation is due to individual traps and
the friction force is $F_{fr} \sim \frac {\tau^{3}}{k\zeta u d} \sim
\frac{N \tau^2}{k d } \frac{u_c}{u}$. It
decreases with the velocity. If $d\ll\frac{\tau^{2}}{k \zeta u}$, as
above, the dissipated energy per trap is of the order of the elastic
energy of the chain section spanning the length $d$ with a tension
$\tau$ and the friction force is $F_{fr}=\tau$.

All these results are summarized on figure \ref{diagramme} in a
velocity-distance between traps $(u,d)$ diagram.

\section{Thermal fluctuations.}
\label{fluc}

In the previous section, we have considered point-like traps and we
have ignored thermal fluctuations describing the chain statistics by
the average position of each monomer. In a more realistic model,
these two assumptions are relaxed. Each trap is an attractive
potential well $V(z-z_{0})$ for the chain end with an extension $b$ around
the trap
position $z_{0}$ and a depth $U_{0}$. The maximum tension that can be
exerted by the
trap $\tau$ is the slope of this potential at the inflection point $z_{1}$. We
want to study here strong trapping and we only focus on the case where
$\frac {\tau b}{k_BT} \gg 1$. This can be achieved for a reasonable
tension if the size of the trap is larger than the monomer size $a$
(typically a few times) but is much smaller than the chain size and
the distance between traps $d$. A typical trapping
potential is sketched on figure \ref{figpotentiel}.a.

Thermal fluctuations have two effects; they can induce a new trapping
of a chain end that has just escaped from the trap; they can also
cause the release of the end monomer even if the tension is lower
than the critical value $\tau$.

In references \cite{wittmer,clement}, Wittmer {\it et al} and Clement {\it  et al} 
have shown that the
readsorption of the chain can
occur until the terminal monomer leaves the first Pincus blob of
size $\xi= k_BT/\tau_{tr}$. The size of this Pincus blob in the strong
trapping limit is smaller than the size of the trap $b$ when the
tension is equal to the critical tension. The retrapping effect can
then be neglected.

The escape of a particle submitted to an external force from  a potential
well can be studied using Kramers rate theory. If the constant
external force is $\tau_{tr}$, particle is submitted to the effective potential 
$U(z)= V(z)-\tau_{tr}z$ (choosing the origin at the center
of the trap). This potential has a minimum value at a position $z_{min}$
corresponding to the
equilibrium position of the particle and a maximum value at a
position $z_{max}$ ($z_{min}<z_{1}<z_{max}$). The escape time due to
thermal fluctuations from
the potential is the first passage time of the particle at the
position $z_{max}$. Within Kramers rate theory \cite{vankampen}, this is given by

\begin{equation}
t_{e}(\tau_{tr}) = t_{1}\exp \left[
\left\{U(z_{max})-U(z_{min})\right\}/k_BT \right]
\label{time1}
\end{equation}

where $t_1$ is given by the curvature of the
potential $U(z)$ at the positions $z_{min}$ and $z_{max}$

\begin{equation}
t_{1} (\tau) = \frac{2\pi
\zeta}{\left(-\frac{d^{2}U}{dz_{max}^{2}}\frac{d^{2}U}{dz_{min}^{2}}\right)^{
1/2}}
\label{time2}
\end{equation}

Our aim here is to use these result to discuss the escape of the trapped
end under the action of the time-dependent tension and 
to calculate the chain tension $\tau_{e}$ when the trapped
end escapes from the potential well. The results of the
previous section can then be used with the effective critical
tension $\tau_{e}$ instead of $\tau$.

In reference \cite{seifert}, two approximations (so-called adiabatic and self-consistent 
force approximation) are proposed to discuss the escape of a particle from a potential 
well under the action of a non constant force. We have checked that both lead to 
very similar results. For a sake of clarity, we present here the simpler self-consistent 
force approximation. In this approximation, the tension is
obtained self-consistently by calculating the escape time from
equation \ref{time1} and \ref{time2} for a tension $\tau_{tr}$ increasing with time and
by imposing that $\tau_{e}=\tau_{tr}(t_{e})$. 

In the limit of small velocities $u\ll u_{c}$, the
chain tension at the end point increases as $\tau_{tr}=\frac {kut}{N}$.
The self-consistency equation for the effective critical tension can
then be written as~:

\begin{equation}
\frac {N\tau_{e}}{ku}= t_{1}\exp \left[
\left\{U(z_{max})-U(z_{min})\right\}/k_BT \right]
\end{equation}

where the potential is calculated with the tension $\tau_{e}$. If the
velocity is very small, the critical tension $\tau_{e}$ is very small
and the escape time can be expanded in powers of the tension. For a
vanishing tension, the escape time is $t_{e}^{0}=t_{1}exp (U_{0}/k_BT)$
where $t_{1} \sim \zeta b/\tau$ (assuming
that the potential has only  one energy scale $U_{0}\sim \tau b$ and
one length scale $b\sim z_{1}$). The effective critical tension is then~:

\begin{equation}
\tau_{e}= \frac{kut_{e}^{0}}{N} \frac 1 {1+\alpha (\tau
b/k_BT)(t_{e}^{0}/t_s(u))}
\label{tension1}
\end{equation}

where $t_s(u)$ is the sticking time introduced in the previous
section and $\alpha$ a numerical constant. This result is valid as long
as the thermal escape time is small enough $t_{e}^{0}\ll
\frac{k_BT}{\tau b} t_s (u)$ leading to
$\frac{u}{u_{c}}\ll \frac{u_{1}}{u_{c}} \sim(Na/b)^{2}\exp -(U_{0}/k_BT)$.

We assume for simplicity that the thermal escape time is larger 
than the Rouse time of the whole chain. The tension is thus uniform along the whole 
chain when the escape from the trap arises;  the velocity $u_{1}$ is then smaller than 
$u_{c}$. In this limit of
vanishing velocities, if the traps are independents, the friction
force is $F_{fr}=\frac {N\tau_{e}^{2}}{2kd}$ and vanishes as $u^{2}$.
If the escape time is smaller than the Rouse time only a fraction of
the chain containing $s \sim (kt_{e}^{0}/\zeta)^{1/2}$ monomers is
stretched. The critical tension is $\tau_{e}= (k\zeta
t_{e}^{0})^{1/2}u$
and the friction force is  $F_{fr}=\frac {s\tau_{e}^{2}}{2kd}$; it
also
vanishes as $u^{2}$.

If the velocity is larger than $u_{1}$, the effective tension increases
between a value of order $k_BT/b$ and $\tau$. The variation of the
effective critical tension explicitly depends on the shape of the
potential and cannot be expressed in simple analytical form. A
simple expression can only be given if the velocity is large enough.
In this case, the two positions $z_{max}$ and $z_{min}$ are close to
the potential inflection point $z_{1}$. One can then expand the
potential around $z=z_{1}$

\begin{equation}
U(z)=U_{1}+ (\tau - \tau_{e}) (z-z_{1}) -\frac 13 \gamma \tau
z_{1}\left(\frac {z-z_{1}}{z_{1}}\right)^{3}
\label{expansion}
\end{equation}

where $\gamma$ is a number of order one.
The effective tension can be calculated from this expansion~:

\begin{equation}
\tau_{e}= \tau \left(1-\frac 1{\beta^{2/3}}
\log^{2/3}\left(t_s (u)\frac{\tau_e}{\tau}(1-\frac
{\tau_{e}}{\tau})^{1/2}/t^0_{1}\right)\right)
\label{tension2}
\end{equation}

where we define the dimensionless number $\beta=(4\tau
z_{1})/(3k_BT\gamma^{1/2})$ and the microscopic time $t^0_{1}=(\zeta
z_{1})/(2 \gamma^{1/2 }\tau)$.
This expansion is valid if $\beta$ is larger than $\log (t_{s}/t_{1})$.
Qualitatively the effective tension is equal to the maximal force
exerted by the potential if the sticking time $t_s$ is smaller
than a time of the order of the thermal escape time.
The effective tension increases smoothly (with a logarithmic dependence in u)
between $k_BT/b$ and $\tau$. At first order we can use the power law found
in the previous section in the framework of the zero temperature theory.

A similar calculation can be made at larger velocities $u\gg u_{c}$.
If the thermal escape time is larger than the Rouse time, the
effective critical tension is close to $\tau$ and a law similar to
eq. \ref{tension2} is found (only the argument of the logarithm is
slightly different). One can thus also use the zero temperature result
for the friction force.

Finally, if the traps are not independent one should compare the
thermal escape time to the sticking time that, in this case, is of
order $d/u$. The escape of the trap is dominated by thermal fluctuations if
the thermal escape time is smaller.

\section{Concluding Remarks}

In this paper, we have studied a very specific model for the friction
of a polymer chain on a solid surface. We have considered a
two-dimensional gaussian polymer chain interacting with traps on a solid
substrate. All the dissipation in this model is due to viscous
friction in the solvent surrounding the chain and there is no
dissipation associated to the substrate. Our main result is that, neglecting 
thermal fluctuations, the traps exert a 
finite friction force on the chain that does
not vanish when the velocity goes to zero. This friction force originates
in the elastic energy stored in the chain which is then dissipated by
viscous friction in the liquid. At higher velocity the friction force
decreases with the velocity due to the fact that the portion of chain
that is stretched gets smaller. If thermal fluctuations are taken
into account, the friction force formally vanishes at zero
velocity. Nevertheless, for most practical cases the traps are strong enough and 
the effect of thermal fluctuations will appear only at infinitely
small velocities. However, a very weak logarithmic dependence of the
friction force on the velocity still remains, due to the fact that the 
trapping cross section depends logarithmically on the velocity.

For a macroscopic sample consisting of a statistical ensemble of 
independent chains, the static friction coefficient is proportional 
to the average number of trapped chains. Furthermore, our model gives an 
interesting result concerning the dependence of  this coefficient with the 
sample history. During a friction experiment at velocity $u$, the average 
fraction of trapped chains is given by $f \sim \frac{t_s (u) u}{d}$. For 
$u \ll u_c$ this fraction is constant and equal to $f_0 \sim \frac{N \tau}{k d}$. 
For $u \gg u_c$ it decreases with the velocity as 
$f_u \sim \frac{N \tau}{k d} \frac{u_c}{u}$. In this case, if the motion 
is stopped fast enough, the static friction coefficient is lower 
than $f_0$ and increases with time until it reaches $f_0$.

Our model may look very simplistic, but we believe that many of the
assumptions made here do not have a strong
influence on the main results although they may change the
quantitative value of the friction force. For example, we have
assumed that only the last monomer of the chain (the chain end that
is not pulled) interacts with the traps. It is easily checked that
identical results are obtained (within a factor $2$) if any monomer
not too close to the chain end interact with the traps. We have
explicitly performed the Rouse calculations in this case, they are
slightly more complicated and we do not present them in this paper.
Another extension that we would like to make in the future is the case
where several monomers of the same chain interact with the traps.
There is of course a possibility that at the same time several
monomers are trapped; this then leads to correlated release (cascades).
We believe however that the same qualitative results would be obtained and that
the friction force would remain finite at low velocities.  We also have
assumed a periodic array of traps. A more realistic approach would be
to introduce disorder both in the positions of the traps and in the
distance between traps. A weak disorder is clearly not important. If
the traps are independent, then the averaging of the dissipation is
straightforward. If the average distance between traps is small, the
tension of the end
monomer remains roughly constant as the tension fluctuations due to
the trappings do not propagate to the pulled monomer. Disorder could
play an important role if it introduces strong fluctuations (such as
rare trapping events with a strong critical tension). It would then be
interesting to study the fluctuations in the friction force.

A last limitation is that we have only studied the stationary
friction state. Transient effects are obviously very important and
must be studied independently.

It is also important to discuss the orders of magnitude of the
predicted effects. Our more quantitative discussion of the release from the
trap clearly shows that the solid-like friction can only be observed if
the energy barrier is large $\tau b/k_BT\gg 1$ otherwise the whole
escape process is thermally activated and there is no solid
friction. This could be achieved for example with a trap size of the
order of a few monomer sizes $b\sim 10a$ and $\tau b/k_BT \sim 10$. The
critical velocity for a chain of $1000$ monomers is then of
the order of $0.1 mm/s$ i.e. in the accessible range for atomic force
microscopy. One must however notice that the chain
tensions can then be rather large and that it is not clear that the
chain elasticity remains gaussian. This would of course change the
quantitative value of the friction force but not the qualitative
physics.

In the future, we would like to use our very simple model to study the
solid friction of a polymer gel moving on a solid substrate in order to
give a quantitative interpretation of the recent experiments of Gong
and Osada \cite{gong}

\newpage

\appendix

\section{Trapped chain.}
\label{A}

It is convenient to perform all calculations using the Laplace transformation,
defined by~:

\begin{equation}
 \tilde{z}(s,p) = \int_0^{+\infty}z(s,t) e^{-pt} dt
\label{laplace}
\end{equation}

For the trapped chain, the following Rouse equation has to be solved:

\begin{equation}
\zeta \left( p \tilde{z}(s,p)-z_{free}\right) = k \frac{\partial^2}{\partial
s^2}\tilde{z}(s,p)
\label{rousetrapped}
\end{equation}

with the boundary conditions~:

\begin{eqnarray}
\nonumber
&&\tilde{z}(s=0,p) = 0 \\
&&\tilde{z}(s=N,p)=\frac{u}{p^2}
\label{limittrapped}
\end{eqnarray}

The solution of this equation is given by:

\begin{eqnarray}
\nonumber
\tilde{z}_{d}(s,p) &=& \frac{u}{p^2}\left[ 1+\frac{1}{2} p T_R \left(
\frac{s}{N} \right)^2 +\frac{\sinh{\sqrt{T_R p}\frac{\left( s - N
\right)}{N}}}{\sinh{\sqrt{T_R p}}} \right] \\
z_{d}(s,t) &=& u T_R \left[ \frac{t}{T_R}\frac{s}{N} + \frac{1}{2} \left(
\frac{s}{N} \right)^2 +\frac{2}{\pi^3}
\sum_{n=1}^{+\infty}\frac{(-1)^n}{n^3} \left( 1 -
e^{-n^2\pi^2\frac{t}{T_R}} \right) \sin{\left( \pi \frac{n}{N}(s-N)
\right)} \right]
\label{soltrapped}
\end{eqnarray}

and the tension along the chain is given by:

\begin{equation}
\tau_{ch}(s,t) = k \frac{u T_R}{N} \left[ \frac{t}{T_R} + \frac sN
+\frac{2}{\pi^2}
\sum_{n=1}^{+\infty}\frac{1}{n^2} \left( 1 - e^{-n^2\pi^2\frac{t}{T_R}}
\right) \cos{\left( \pi \frac{n}{N}(s-N) \right)} \right]
\label{tens}
\end{equation}

At large time $t \gg T_R$, the tension on the trapped monomer behaves
as~:

\begin{equation}
\tau_{tr}(s=0,t) \simeq \frac kN u t
\label{tenstrappedshortt}
\end{equation}

which corresponds to the tension of N springs under the extension $ut$.

 For short times
$t \ll T_R$, the tension on the trapped monomer behaves as~:

\begin{equation}
\tau_{tr}(s=0,t)=\frac{2}{\sqrt{\pi}} \frac kN u T_R \sqrt{\frac{t}{T_R}}
\label{tenstrappedlarget}
\end{equation}
As given in the main text, the sticking time is obtained when
this tension is equal to the critical value $\tau$.

In this case, the number of extended springs is $s_0(t) \sim N
\sqrt{\frac{t}{T_R}}$.

\newpage

\section{Detachment}
\label{B}

After the detachment of the trapped monomer, the chain relaxes to the
steady state, and its
elastic energy is dissipated. The Rouse equation becomes:

\begin{equation}
\zeta \left( p \tilde{z}(s,p)-z_{d}(s)\right) = k \frac{\partial^2}{\partial
s^2}\tilde{z}(s,p)
\label{rouserelaxed}
\end{equation}

where $z_{d}(s) \simeq u T_R \left[ \frac 12 \left( \frac{s}{N} \right)^2 +
\frac{t_s}{T_R}
- 4\frac{t_s}{T_R}i^2erfc \left( \frac 12 \sqrt{\frac{T_R}{t_s}\frac sN}
\right) \right]$ ($i^nerfc(x)$ is a repeated integral of the
error function defined in \cite{abramovitz}) gives
the conformation of the chain at the time where the chain
detaches form the trap . The boundary conditions are:

\begin{eqnarray}
\nonumber
\frac{\partial \tilde{z}}{\partial s}(s=0,p) &=&  0 \\
\tilde{z}(s=N,p) &=& u T_R \left(\frac 12  + \frac{t_s}{T_R} \right)\frac
1p + \frac{u}{p^2}
\label{limitrelaxed}
\end{eqnarray}

The solution of (\ref{rouserelaxed}) is given by:

\begin{equation}
\tilde{z}(s,p) = A u_+(s,p) + B u_-(s,p) \frac{1}{2N} \sqrt{\frac{T_R}{p}}
\left( u_-(s,p) \int_0^s z_d(s') e^{\sqrt{p T_R}\frac{s'}{N}}ds' -
u_+(s,p) \int_0^s z_d(s') e^{-\sqrt{p T_R}\frac{s'}{N}}ds'\right)
\label{solrelaxed}
\end{equation}

where $u_\pm = e^{\pm \sqrt{p T_R}\frac{s}{N}}$, $A$ and $B$ are two
constants determined
by (\ref{limitrelaxed}). The first boundary condition gives
$A=B=\frac{\alpha}{2}$. In the limit of large times,
 the second condition gives $\alpha \simeq
\frac{\tilde{z}(N)}{\cosh(\sqrt{p T_R})}$.
At large times, the tension on the pulled monomer is given by:

\begin{eqnarray}
\nonumber
\tilde{\tau_{d}}(N,p) &\simeq& N \zeta \frac{u}{p} + 8 i^3erfc(0) N \zeta u
\sqrt{\frac{t_s}{T_R}} t_s\\
\tau_{d}(N,t) &\simeq& N \zeta u + 8 i^3 erfc(0) N \zeta u
\sqrt{\frac{t_s}{T_R}}
t_s \delta(t)
\label{tensrelaxed}
\end{eqnarray}
By integrating this tension over time one finds the dissipated energy
given in the main text (equation \ref{work}).

\newpage

\section{Hydrodynamic interactions.}
\label{C}

The calculations presented in the main text of this paper have been
made using the Rouse dynamics for the polymer chains on the solid
surface. This ignores two effects~: the direct friction of the chains
on the solid surface and the hydrodynamic interactions between
monomers. The hydrodynamic interactions close to a surface are
difficult to calculate explicitly. However if we remain at a scaling
level, it is sufficient to introduce the local stretching blobs
\cite{pincus}. If
the local tension of the chain is $\tau_{ch}$, the local blob has a
size $\xi_{t}\sim k_BT/\tau_{ch}$ and contains $g_{t}\sim
(\xi_{t}/a)^{2}$ monomers. The friction force on a blob is $6 \pi \eta
\xi_{t}$ and the hydrodynamic interactions are screened at the scale of
the blob so that the dynamics of the chain of blobs can by studied
using the Rouse model; the total friction is the sum of the frictions
over all the blobs.
As in the Rouse model, a free chain pulled by one end with a
velocity $u$ has on average
a trumpet like shape studied in details by Brochard and her
coworkers \cite{brochard}. The Zimm equation of motion is obtained by
writing a force balance on the blobs

\begin{equation}
k \frac{\partial^{2}z}{\partial s^{2}}= 6\pi \eta u \frac {\partial
z}{\partial s}
\label {zimm}
\end{equation}

The average position of monomer $s$ is then
\begin{equation}
z_{free} (s)= \left(\frac{k_{B}T}{6\pi \eta u}\right)^{1/2} \left[\exp
(\frac{6\pi \eta u s}{k}) -1 \right]
\end{equation}

The force balance on a section of chain containing $s$ monomers close
to the free end point is obtained by integration of the equation of
motion
$k\frac{\partial z}{\partial s}= 6\pi  \eta u z$. Note that the
tension increases very rapidly from the free end point and that the
nonlinear elasticity certainly becomes relevant. We do not discuss
this effect here and refer to reference \cite {brochard}.

For a chain trapped by one end on the surface,
the sticking time does not depend on the dynamical model at low
velocities (if it is larger than the characteristic relaxation time of
the chain which is the Rouse time of the chain of blobs) $t_s =
N\tau/(uk)$. The dissipated energy is still the elastic
energy of a chain with uniform tension $\tau$.

At high velocities, the sticking time is obtained by a scaling
argument similar to that used in the Rouse model. When the end point
escapes from the trap after a time $t_s$, a chain section of $s(t_s)$
monomers is stretched with a tension $\tau$. The characteristic
relaxation time of this chain section is the Rouse time of the
corresponding chain of blobs $t_s =\frac{6\pi \eta s^{2}\tau}{k^2}$.
Using the
integrated equation of motion, we find the sticking time $t_s =
\frac{\tau}{6\pi \eta u^{2}}$. It has the same dependence on the
velocity as in the Rouse model but a different dependence on the
tension which is due to the fact that one must  now consider the chain
as a Rouse chain of Pincus blobs and not of monomers.

The crossover velocity is obtained by comparing the low and high
velocity results $u_{c}= \frac {k}{6\pi \eta N}$ and the energy
dissipated on a trap at high velocity is the elastic energy of a chain
section containing $s(t_s)$ monomers: $W_{d}= \frac{\tau^{2}}{6\pi \eta
u}$. Here also we find the same dependence on velocity but a different
tension dependence if we compare to the Rouse model.

 \newpage

\section{Trapping cross section}
\label{D}

In this appendix, we study the trapping cross section for the free
chain end by the potential well of a trap on the surface. We focus on 
the undisturbed part of chain end of size $\delta r$. At very low velocity $u \ll
u_{0}=R/T_{R}$ ($T_{R}$ is the Rouse time) and $\delta r$ is the chain radius
$R$. At
higher velocity the chain is stretched by the hydrodynamic friction and
$\delta r$
is equal to the local stretching blob size ($\delta r \sim u^{-1/3}$). If we
take the origin at the center of the first blob containing the free chain end, the 
surface monomer concentration at a distance $r$ is~:

\begin{equation}
c(r) \sim \delta r^{-2} \exp-(r/\delta r)^{2}
\label{conc}
\end{equation}

\begin{figure}
\vskip2truecm
\centerline{\epsfxsize=10truecm \epsfbox{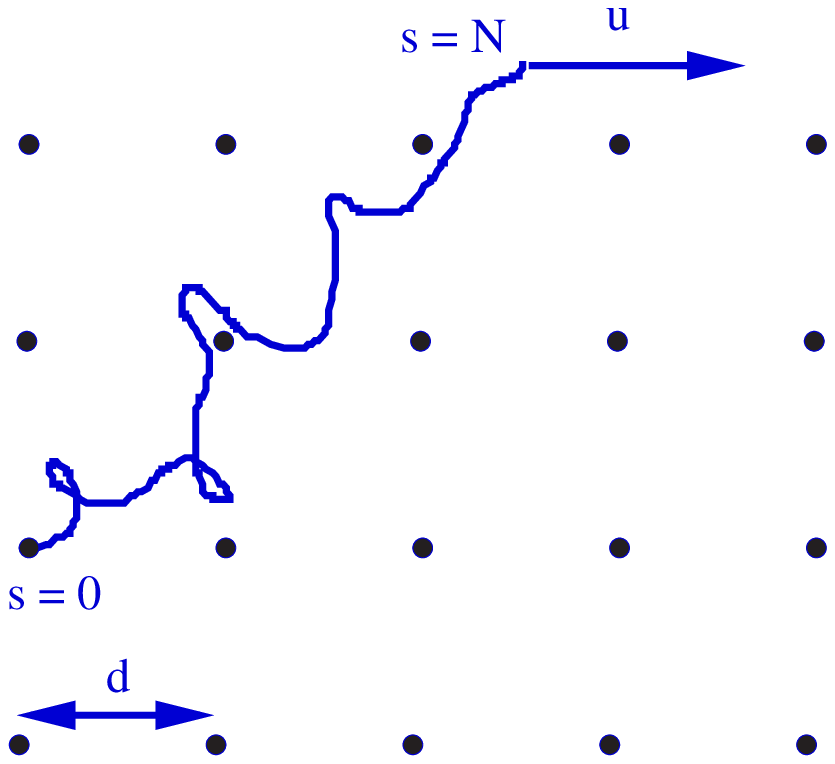}}
\caption{}
\label{generale}
\vskip1truecm
\end{figure}

\begin{figure}
\vskip2truecm
\centerline{\epsfxsize=10truecm \epsfbox{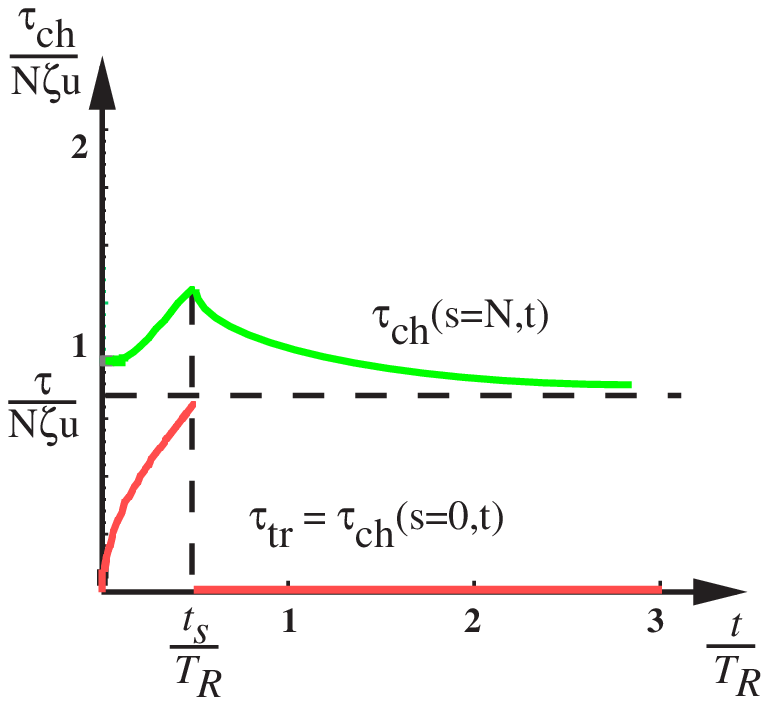}}
\caption{}
\label{tension}
\vskip1truecm
\end{figure}

\begin{figure}
\vskip2truecm
\centerline{ \epsfxsize=10truecm \epsfbox{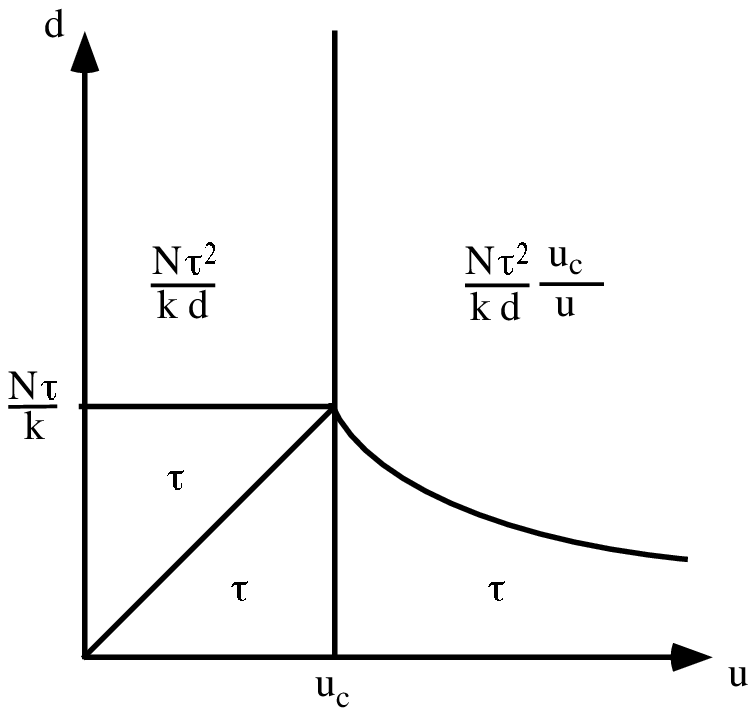}}
\caption{}
\label{diagramme}
\vskip1truecm
\end{figure}

\begin{figure}
\centerline{ \epsfxsize=10truecm \epsfbox{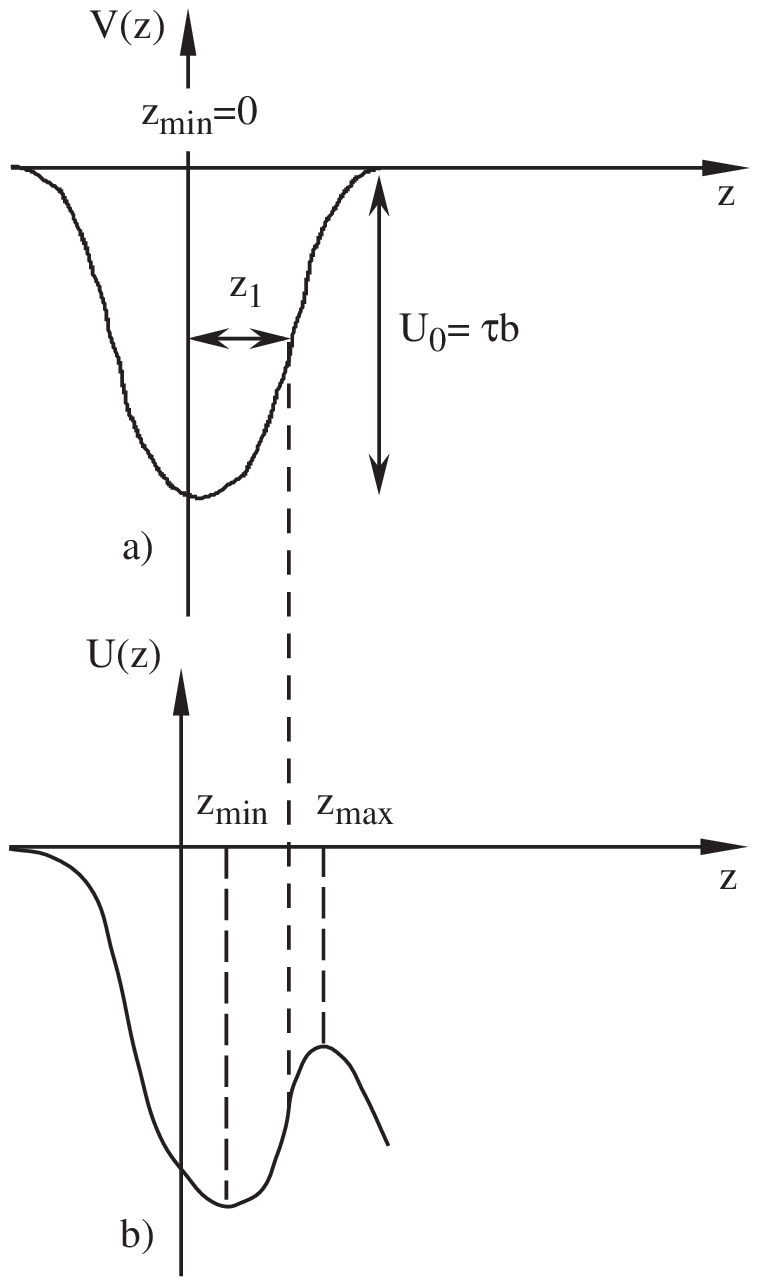}}
\caption{}
\label{figpotentiel}
\end{figure}

The trap acts as an absorbing site and we consider that the chain end
gets trapped when it diffuses to the surface of the trap of radius $b$
(figure \ref{generale}). If the trap is at a distance $r$ from the average 
position of the end point, the concentration gradient close to the 
trap is $c(r)/b$. The total flux of end point towards the trap is thus given by~:

\begin{equation}
q=2\pi b D c(r)/b
\end{equation}

The relevant diffusion constant D is equal to the
Rouse diffusion
constant of the whole chain $D=k_BT/(N\zeta)$ at very low velocity and to
the Rouse
diffusion coefficient of the stretching blob $D=k_BT/(s\zeta)$ ($s$ is the
number of monomers in the blob) at high velocity. Note that in
two dimensions there are logarithmic corrections to this result as there is no
stationary diffusion state. If the impact parameter (the minimum distance
between
the trap and the average chain end position) is $c$ and if the
distance is minimum at time $t=0$ the distance $r$ varies with time as
$r^{2}=c^{2}+u^{2}t^{2}$. The probability $p(c)$ that the chain with an
impact parameter $c$ gets trapped is obtained by integration over time
of the flux $q$, $p(c)\sim \frac {u_{0}}{u} \exp-(c/R)^{2}$. This
probability is of course smaller than one. The trapping
probability is thus equal to one if $c<R \log^{\frac 12} (u_{0}/u)$ and decays
as a gaussian for larger values of $c$. The trapping cross section (a length in 
two dimensions) is obtained by integration over
the impact parameter
\begin{equation}
\sigma \sim R \log^{\frac 12} (u_{0}/u)
\label{cross}
\end{equation}

At higher velocity $u>u_{0}$, the trapping probability is always
smaller than $1$ and one can integrate the flux $q$ to find the
trapping probability and then the cross-section. One must however take
into account that $\delta r$ is now the size of the stretching blob
and that the diffusion constant is not the diffusion constant of the
whole chain but the one of a chain of size $\delta r$. The result is
\begin{equation}
\sigma \sim  \left(\frac {a^{2}k_BT}{\zeta u} \right)^{1/3}
\end{equation}

The average distance between trapping events can then be calculated
from the surface density of traps $n$, $d= 1/\sigma n$.

\newpage

\centerline{{\large {\bf Figures caption}}}

\vspace{2cm}

\begin{itemize}

\item{{\bf Figure 1:} Polymer chain pulled at constant velocity $u$ on a periodic 
2D lattice of traps.}

\item{{\bf Figure 2:} Force (normalised by N$\zeta u$) applied on the trapped 
monomer (s=0, black) and on the pulled one
(s=N, grey) versus time (normalised by T$_R$), according to equation 
\ref{tenstrapped}.}

\item{{\bf Figure 3:} Friction force in a ($u$, $d$) diagram for the many traps problem at 
zero temperature.}

\item{{\bf Figure 4:} Trapping potential a) $V(z)$ and b) $U(z) = V(z)-\tau_{tr}z$.}

\end{itemize}

\newpage


\end{document}